\documentclass[aps,prl,twocolumn,showpacs,longbibliography]{revtex4-1}

\usepackage{graphicx}
\usepackage[colorlinks,linkcolor=red,citecolor=blue]{hyperref}
\usepackage{breakurl}
\usepackage[ansinew]{inputenc}
\usepackage{array}
\usepackage{color}
\usepackage{amsmath}
\usepackage{amsxtra}
\usepackage{amstext}
\usepackage{amssymb}
\usepackage{physics}
\usepackage{latexsym}
\usepackage{float}
\usepackage{multibib}
\newcites{supp}{Supplementary References}
\definecolor{darkgreen}{rgb}{0,0.5,0}
\definecolor{darkblue}{rgb}{0,0,0.7}

\begin{document} 

\title{Observation of coherent oscillations in association of dimers from a thermal gas of ultracold atoms.}

\author{Roy~Elbaz$^{1}$}
\author{Yaakov~Yudkin$^{1}$}
\author{P.~Giannakeas$^{2}$}
\author{Jan-Michael Rost$^{2}$}
\author{Chris~H.~Greene$^{3,4}$}
\author{Lev~Khaykovich$^{1}$}

\affiliation{$^{1}$Department of Physics, QUEST Center and Institute of Nanotechnology and Advanced Materials, Bar-Ilan University, Ramat-Gan 5290002, Israel}
\affiliation{$^{2}$Max Planck Institute for the Physics of Complex Systems, N\"othnitzer Strasse 38, 01187 Dresden, Germany}
\affiliation{$^{3}$Department of Physics and Astronomy, Purdue University, West Lafayette, Indiana 47907, USA }
\affiliation{$^{4}$Purdue Quantum Science and Engineering Institute, Purdue University, West Lafayette, Indiana 47907}%

\date{\today}

\begin{abstract}
We report the observation of coherent oscillations in conversion efficiency of weakly-bound dimers formed from a thermal gas of ultracold atoms.
Finite thermal energy of the gas causes loss of coherence when a broad continuum is resonantly coupled to a discrete bound state.
Restoration of the coherence can be achieved through non-adiabatic transitions of the dressed molecular energy level that are induced by a strong modulation pulse with fast envelope dynamics. 
Conditions to observe coherent oscillations are verified, and control of their properties is demonstrated.
The main experimental findings are supported by theoretical modeling and numerical calculations.
The observed results may lead to a renewed interest in general studies of a discrete energy level coupled to a broadband continuum when the properties of both are fully controlled.
\end{abstract}

\maketitle

When coupling between a discrete energy level and a broad continuum is introduced, dynamics is expected to be fully incoherent due to vanishingly short coherent memory of the latter.
Numerous quantum electrodynamics systems such as spontaneous emission of an excited atom, photoionization of an atom and photodissociation of molecules are all fit within the framework of this general scheme~\cite{CohenTannoudji}.
A similar system can be realized with ultracold atoms in the vicinity of collisional resonances which are characterized by the existence of a weakly bound dimer state.
An original experimental technique to probe this state was demonstrated in Ref.~\cite{Thompson05} which consisted of association of dimers from a thermal gas of atoms by modulating the magnetic field bias at a frequency that matches the binding energy.
Surprisingly, coherent Rabi-type oscillations in the dimers conversion efficiency has been reported and remained unexplained since then despite the number of theoretical approaches devoted to the subject~\cite{Hanna07,Bazak12,Brouard15,Langmack15,Mohapatra15} and despite the straightforward analogy with the mentioned above systems subject to broadband decoherence mechanisms.

A recent theoretical proposal suggests a mechanism that leads to the restoration of coherence in such a system~\cite{Giannakeas19}.
Here we carefully explore its conditions and provide the experimental verification.
Our studies reveal the crucial importance of fast dynamics initiated by the excitation pulse envelope. 
The mechanism bares similarities with St\"uckelberg oscillations demonstrated earlier in multiphoton ionization~\cite{Jones95} and in Rydberg excitation~\cite{Bengs22} of atoms.
More recently, similar effect has been suggested theoretically also for x-ray ionization by intense femtosecond  laser pulses where their steep raise and  fall  induces dynamic interference in the photoionization spectra ~\cite{Demekhin12a,toyotanjp2015,Baghery17,Ning18,com2}.

 \begin{figure}
	\centering
	\includegraphics[width=1.05\columnwidth]{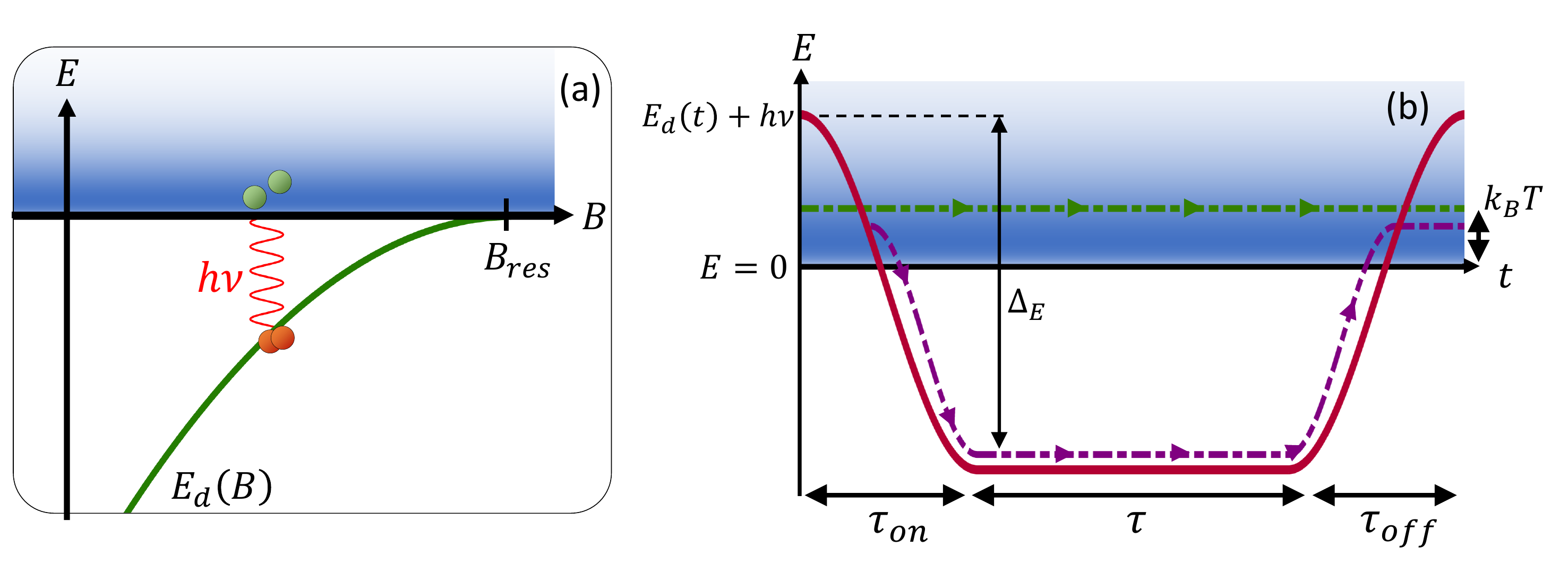}
	\caption{\label{fig:Scheme}
		\textbf{Schematics of physical mechanisms.}
		(a) - Universal dimer energy level in the vicinity of a Feschbach resonance. The binding energy decreases when the magnetic field increases to suit the specifics of the Feshbach resonance used in the experiment. (b) - Principles of the interferometer: when the interaction is switched on, fast non-adiabatic transition of the dressed energy level through a thermal atomic ensemble ($\tau_{\rm on,off} \ll \tau_{\rm th}=h/k_B T$) splits the wave function into two pathways: one follows the atomic continuum (green dashed-dotted line) while another one follows the bound state shifted below the threshold by strong coupling ($\Delta_E \gg k_B T$). When interaction is switched off the wave function recombines and the outcome depends on the accumulated phase difference between the paths.
	}
\end{figure}

Conversion of ultracold atoms into a weakly bound dimer 
is executed by modulation of the magnetic field bias $B(t)$ with a resonance frequency $\nu$, as shown in Fig.~\ref{fig:Scheme}(a).
To illustrate the physical mechanism behind the coherent oscillations in the molecule conversion efficiency, the induced interaction dynamics is represented in Fig.~\ref{fig:Scheme}(b) as a dressed state energy level in the rotating wave approximation.
Before the modulation pulse is switched on, the dressed energy level $E_\textrm{d}+h\nu$ is degenerate with the thermal continuum.
When the modulation strength grows, $E_d$ experiences repulsion from the continuum and is pushed down to lower energies.
Ultimately, at strong enough modulation $E_d+h\nu$ crosses the dissociation threshold and opens a gap with the continuum  ($\Delta_E \gg k_B T$, see Fig.~\ref{fig:Scheme}(b)).
If the dynamical changes ($\tau_{\rm on,off}$) of the dressed level are fast compared to timescale set by the thermal energy ($\tau_{\rm th}=h/k_B T$) the nonadiabatic coupling creates an effective beam-splitter each time the threshold is passed and forms a two-path interferometer for a wave function to either make a dimer or remain in an unbound state (see Fig.~\ref{fig:Scheme}(b)).
As was demonstrated in Ref.~\cite{Giannakeas19} such an interferometer should be robust against thermal averaging and, as we show here, it is responsible for the observed coherent oscillations.

Our study settles a long-term quest for a deeper understanding of an experimental technique routinely used for nearly two decades in many species and mixtures such as homo-~\cite{Papp06,Lange09,Gross11,Dyke13} and hetero-nuclear atomic systems~\cite{Gaebler07,Weber08,Wang15} and in strongly magnetic atoms~\cite{Frisch15,Ye22}.
Revealing a non-adiabatic effect related to the fast dynamics of the modulation pulse envelope provides an old problem with a fresh perspective and novel and unexpected relations with other systems such as, for example, photoionization with intense femtosecond laser pulses ~\cite{Demekhin12a,toyotanjp2015,Baghery17,Ning18,com2}.
In this context, our work complements and enhances a recently established quantum simulator of ultrafast phenomena with ultracold atoms~\cite{Senaratne18}.
Even in a broader context, our results pave the way for studies of a discrete energy level coupled strongly to a broadband continuum in a well controlled environment provided by ultracold atoms in which both the energy level and the continuum properties can be adjusted.


The experiment is performed in a thermal gas of $^7$Li atoms, polarized in the $|m_{J}=-1/2, m_{F}=0\rangle$ state and evaporatively cooled to a temperature of $T\approx1.5\;\mu$K in a crossed-beam optical trap~\cite{Gross08}.
Feshbach dimers are associated in the vicinity of the $893.78$G $s$-wave Feshbach resonance by means of a magnetic field modulation at a frequency of $\nu=2.538$~MHz. 
The latter is in resonance with the dimer's binding energy when the magnetic field bias is set to $884.94$G, which corresponds to the scattering length $a=425a_0$~\cite{Gross11,Julienne14}.
The modulation is applied by 2 auxiliary coils on both sides of the vacuum chamber, aligned with the magnetic field bias.

In the first set of experiments we explore the effect of large modulation amplitudes $b_0$ on the dimer binding energy. 
Up to a second order in perturbation theory the dimer energy (averaged over a modulation period) is expressed as a dressed state~\cite{supMat}:
\begin{equation}
	E_{d}(b_0)\approx A E_\textrm{bare}\left[1+\frac{3}{2A}\left(\frac{ b_0}{B_\textrm{bare}-B_\textrm{res}}\right)^2\right],
	\label{eq:Eshift}
\end{equation}
where $E_\textrm{bare}=-\hbar^2/ma_\textrm{bare}^2$ is the dimer binding energy for vanishing modulation strength in the universal limit, $a_\textrm{bare}=|a_{bg}\Delta/(B_\textrm{bare}-B_\textrm{res})|$ and $B_\textrm{bare}$ are the corresponding scattering length and magnetic field, $\Delta$ and $B_\textrm{res}$ are the width and location of the Feshbach resonance, and $A\approx 0.86$ corrects $E_\textrm{bare}$ for the finite range of the potential and the resonance strength.
When the modulation frequency $\nu$ satisfies the resonance condition, i.e. $E_d=h\nu$, dimers are formed most efficiently~\cite{Thompson05}.
Thus, characterizing the optimal dimer conversion efficiency as a function of modulation strength and using Eq.~(\ref{eq:Eshift}) allows for accurate calibration of $b_0$ at the position of atoms. 
This information is important to verify the first condition to observe St\"uckelberg oscillations.
 
 \begin{figure}
	\centering
	\includegraphics[width=0.95\columnwidth]{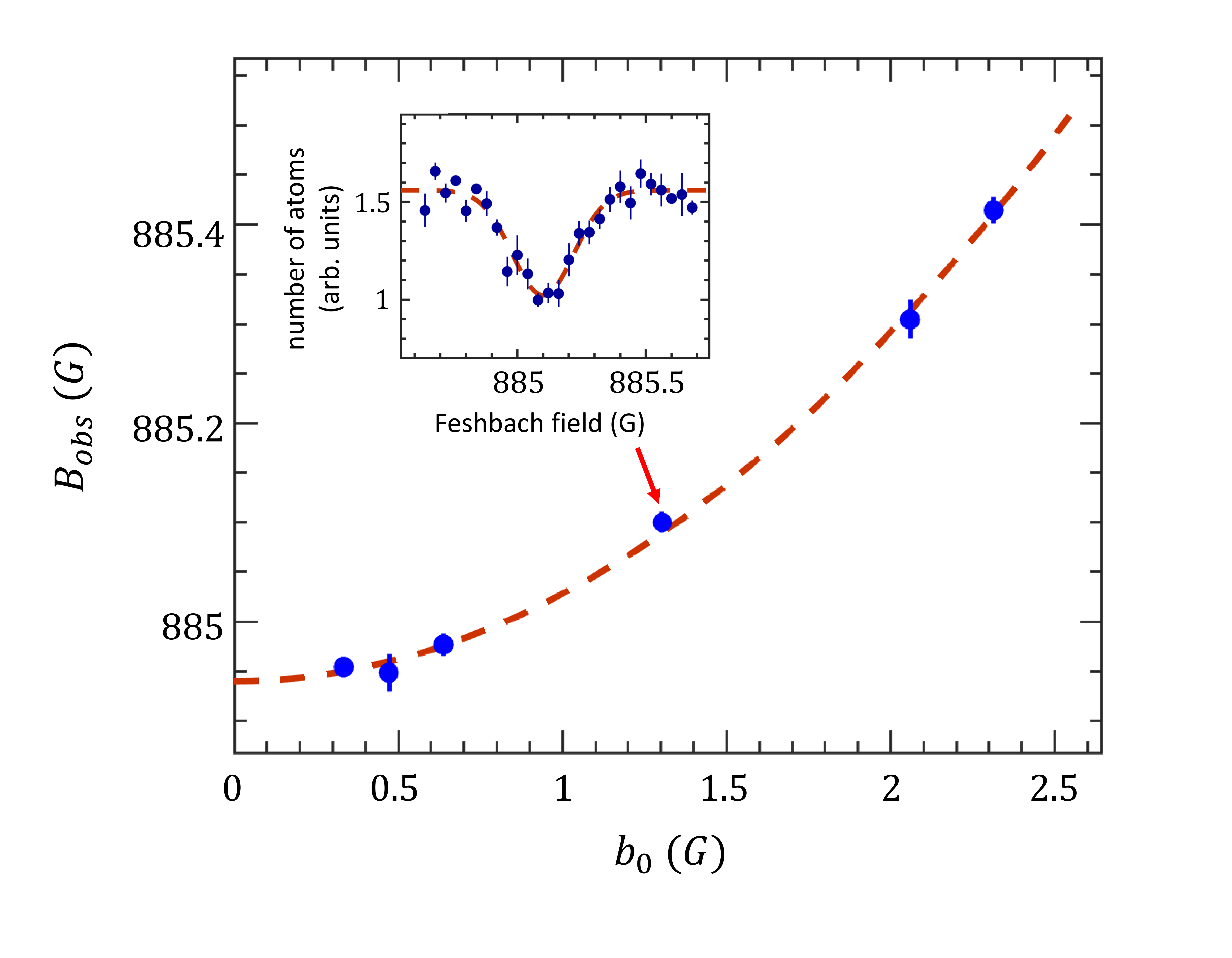}
	\caption{\label{fig:resShift}
		\textbf{Shift in resonance.}
		Optimal bias field for dimer association $B_{\text{obs}}$ as a function of the magnetic field modulation amplitude $b_0$. Each blue data point stands for a dimer association spectroscopy loss measurement. The dashed red curve is drawn according to Eq.~(\ref{eq:Bshift})~\cite{supMat}. In the inset an example for such a spectroscopy measurement is shown, where the remaining number of atoms in the trap after the rf pulse of a fixed frequency $\nu $ is plotted versus the bias magnetic field. 
	}
\end{figure}

In the experiment the modulation frequency is fixed and we scan the magnetic field bias to find the resonance response in dimer association.
Recasting Eq.~(\ref{eq:Eshift}) in this language we get:
\begin{equation}
\begin{split}
	&|B_{\text{obs}}-B_\textrm{res}|\approx \\ &\sqrt{A} |B_\textrm{bare}-B_\textrm{res}|\left[1+\frac{3}{4A}\left(\frac{b_0}{B_\textrm{bare}-B_\textrm{res}}\right)^2\right].
\end{split}
\label{eq:Bshift}
\end{equation}
For each value of $b_0$ we scan the magnetic field bias and measure the number of remaining atoms in the trap after a short waiting time during which the molecules decay from the trap due to inelastic collisions (see Fig.~\ref{fig:resShift} (inset)). 
The pulse duration is varied from $150\mu$s to $500\mu$s to keep the loss contrast ($(N_{\rm bg}-N_{\rm min})/N_{\rm bg}$, where $N_{\rm bg,min}$ is the background and the minimal number of atoms) roughly constant and $\sim 40\%$. 
The resulting loss resonance is fitted with a Gaussian profile and the center value $B_{\text{obs}}$ is extracted.
In Fig.~\ref{fig:resShift} $B_{\text{obs}}$ is shown as a function of $b_0$ revealing that a maximal shift of $B_{\text{obs}}-B_\textrm{bare}=466$~mG can be readily obtained with a modulation amplitude as large as $b_0=2.3$~G at the position of atoms~\cite{supMat}.
This corresponds to an energy shift of $\Delta_E\approx h\times260$~kHz which surpasses the thermal energy by almost one order of magnitude and satisfies the condition to observe St\"uckelberg oscillations.
Note that having this large energy margin is essential for the success of the experimental demonstration as will be clarified below.

\begin{figure}
	\centering
	\includegraphics[width=1.0\columnwidth]{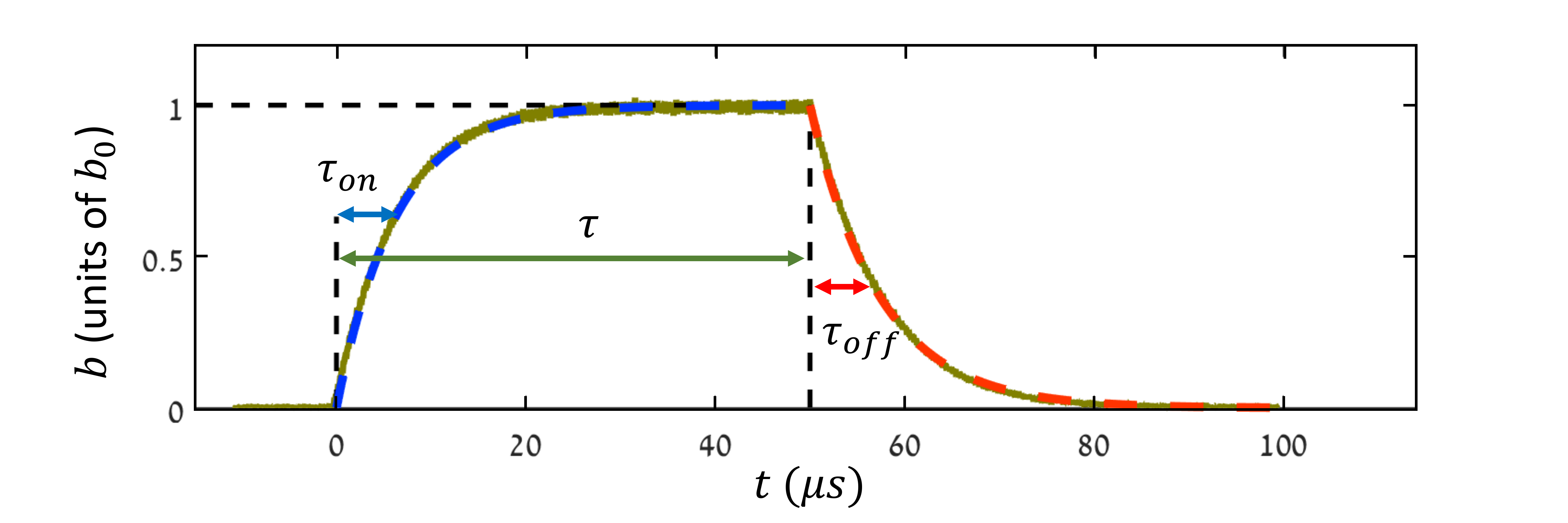}
	\caption{\label{fig:pulseForm}
		\textbf{Modulation pulse envelope.}
		The magnetic field modulation envelope is shown together with fitting to Eq.~(\ref{eq:pulseEnv}). The pulse is turned on at $t=0$ for a duration $\tau$ (blue dashed line), and is turned off at $t=\tau$ (red dashed line). The rise and fall times $\tau_{\text{on,off}}$ are marked by blue and red double arrows respectively. The maximal pulse amplitude, denoted by $b_0$, is reached for $\tau\gg\tau_{\text{on}}$.
	}
\end{figure}

Consider next the main experimental protocol used for the observation of the St\"uckelberg oscillations.
The modulation coils are part of a resonant RLC circuit to maximize the available modulation amplitude. 
This results into a time-dependent magnetic field of the following form:
\begin{equation}
    B(t)=B+b(t)\sin(2\pi \nu t),
    \label{eq:tdmagFiel}
\end{equation}
where $b(t)$ is the pulse envelope described by the relation:
\begin{equation}
b(t)=
\begin{cases}
b_0\left(1-\text{exp}(-t/\tau_{\text{on}})\right), & 0\leq t<\tau\\
b_1\text{exp}(-(t-\tau)/\tau_{\text{off}}), & t\geq\tau
\end{cases}
\label{eq:pulseEnv}
\end{equation}
where $\tau$ is the pulse duration, $b_1=b_0\left(1-\text{exp}(-\tau/\tau_{\text{on}})\right)$ is the amplitude at the moment the pulse is switched off and
$\tau_{\text{on/off}}$ are the rise and fall times of the pulse envelope.
Fig.~\ref{fig:pulseForm} shows the pulse's envelope as measured by a pick-up antenna.
By fitting it to Eq.~(\ref{eq:pulseEnv}) the rise and fall times are extracted to be $\tau_{\text{on}}=6.27~\mu$s and $\tau_{\text{off}}=7.38~\mu$s. 
Note that the typical timescale of the system attributed to the finite thermal energy is longer than the envelope's dynamics ($\tau_{\text{th}}\approx30~\mu$s $> \tau_{\text{on,off}}$) satisfying the non-adiabaticity criterion (see Fig.~\ref{fig:Scheme}(b)).
The dressed energy level dynamics $E_d(t)$ is now described by Eq.~(\ref{eq:Eshift}) where $b_0$ is substituted with $b(t)$.

\begin{figure*}
	\centering
	\includegraphics[width=1.\textwidth]{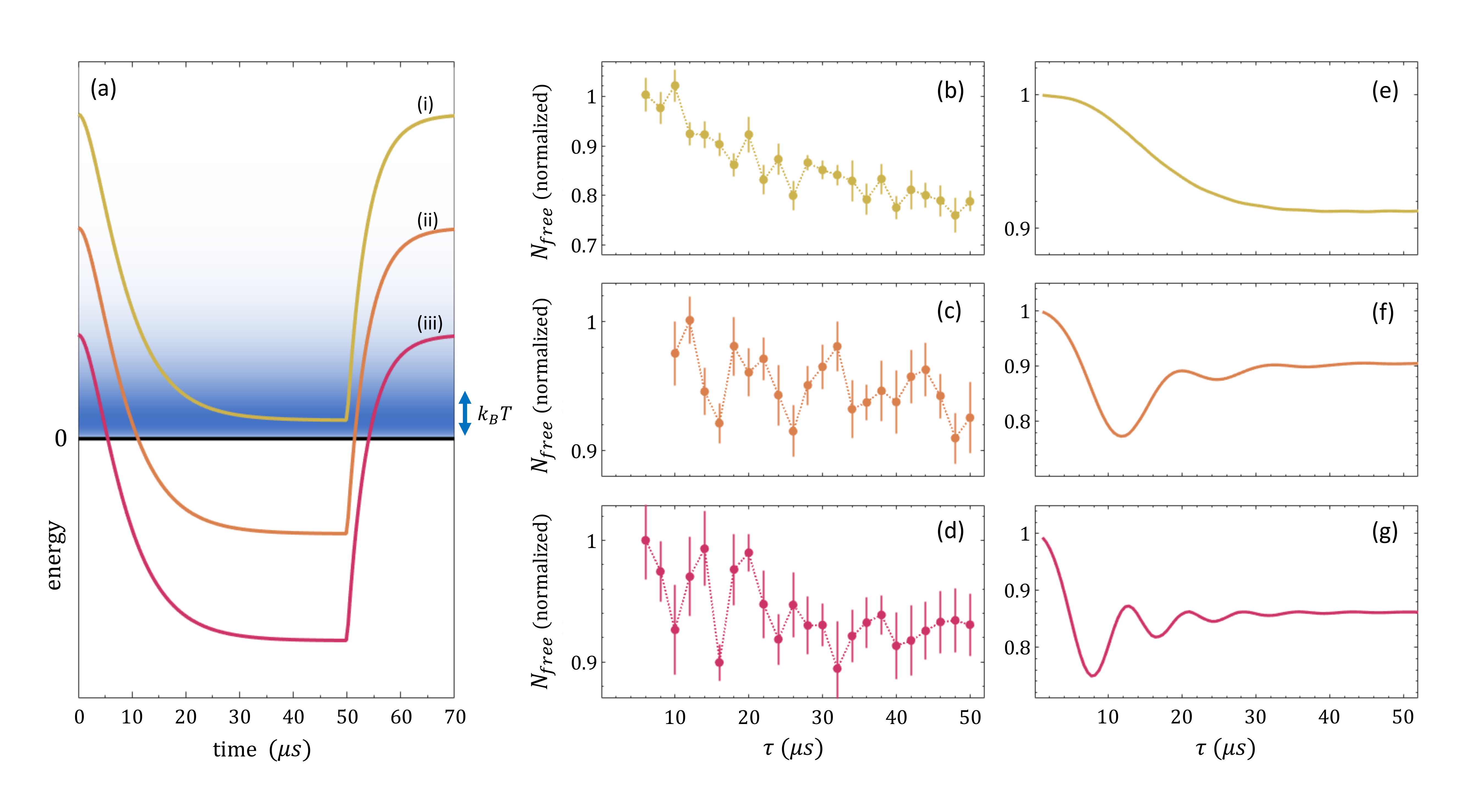}
	\caption{\label{fig:exp}
		\textbf{Experimental results and numerical calculations.} (\textbf{a}) Time evolution of the dressed dimer energy level $E_d(t)+h\nu$ as the modulation pulse is applied. The initial energy at $t=0$ is different for the three scenarios. The thermal population of free atoms is gradiated by the blue color gradient with an energy width of $k_B T$. Cases (i), (ii) and (iii) are marked in yellow, orange and red solid lines respectively.
		(\textbf{b}) shows the remaining number of free atoms in the trap for different modulation pulse durations $\tau$ corresponding to the case (i) in which the dimer is in resonance with the continuum when the pulse is at its maximal modulation amplitude.
		(\textbf{c}), (\textbf{d}) show the same plot for cases (ii) and (iii) respectively where St\"uckelberg oscillations are visible. 
		(\textbf{e})-(\textbf{g}) show corresponding numerical calculations as per Ref.~\cite{Giannakeas19} with parameters suitable for our experimental realization.
	}
\end{figure*}

Before the modulation pulse is turned on, a bias field $B$ is applied, such that $h\nu>|E_\textrm{bare}|$.
As the pulse is turned on $E_d(t)$ is shifted to lower values, and when the pulse is turned off the energy level goes back to its initial value. 
We explore three different scenarios illustrated in Fig.~\ref{fig:exp}(a). 
They differ from each other only by the applied bias field $B^{(i)}$ ($i\in\{\text{i,ii,iii}\}$), so that although the initial energies $E_d(t=0)$ are different, the modulation amplitude $b_0$ remains the same for all scenarios. 
This approach assures minimal changes in the experimental conditions which is less so if, for example, the pulse intensity is varied instead.
To keep the validity of nonadiabaticity in all scenarios we start above the threshold with at least $\sim2 k_B T$ (see Fig.~\ref{fig:exp}(a)).
This forces the dressed dimer energy level to pass the majority of populated continuum rapidly, and is especially relevant to avoid the problematic long tail of weak modulation when the pulse is switched off (see the red dashed tail in Fig.~\ref{fig:pulseForm}).

For the scenarios (ii) and (iii) we expect St\"uckelberg oscillations in dimer conversion efficiency.
When the pulse amplitude starts to grow the initial non-adiabatic passage of the dressed energy level through the continuum imitates the entrance beam splitter of an interferometer at which the initial wave function splits between the free atoms and the bound state pathways.
As the pulse amplitude continues to grow, the two paths decouple and accumulate a dynamic phase difference according to their energy separation.
When the pulse is turned off the dressed dimer level crosses the continuum again, this time imitating the exit beam splitter at which the two pathways of the wave function recombine.
The resulting interference is measured through counting the number of free atoms as a function of pulse duration $\tau$ (see Fig.~\ref{fig:pulseForm}).
Note that the population of the dimer state is hidden in the dark because, prior to detection, the magnetic field bias is shifted to deepen the dimer binding energy to $\sim6$~MHz.
Thus, the bound state remains decoupled from the weak, on-resonance detection light.
In the first scenario (i) of Fig.~\ref{fig:exp}(a) the dynamics is mostly adiabatic for energies near $k_BT$ which efficiently eliminates the entrance beam splitter.
Although the exit beam splitter is still created by the sharp end of the pulse, St\"uckelberg oscillations are not expected in this case.

The experimental results and the corresponding numerical calculations based on the integration of the time dependent Schr\"odinger equation are shown in Fig.~\ref{fig:exp}(b,c,d) and (e,f,g) respectively.
Each data point is an average of $16-32$ runs, and the error bars signify the standard deviation. 
All measurements are performed with the maximal modulation amplitude of $b_0=2.3$~G.
The numerics are done according to prescriptions of Ref.~\cite{Giannakeas19} with parameters suitable for our experimental realization and they show qualitative agreement with the measurements (for details see \cite{supMat}).
Quantitative discrepancies are discussed at the end of this section.

In case (i), shown in Fig.~\ref{fig:exp}(b), the bias field is set to $B^{(\text{i})}=B_{\text{obs}}(b_0=2.3$G$)$ (see Fig.~\ref{fig:resShift}) and the dressed level becomes resonant with the continuum only when the pulse amplitude reaches its maximum. 
The interferometric sequence is not realized in this case which can be identified in the monotonic decrease in number of atoms as a function of pulse duration $\tau$. 
For $\tau<10\;\mu$s and $\tau>40\;\mu$s, the signal agrees with static saturation.
In the former case the pulse amplitude is still weak and the modulation frequency is off-resonance as can be seen in Fig.~\ref{fig:exp}(a), keeping the production efficiency low.
The latter is obtained when the rates of association and dissociation of dimers due to on-resonance modulation become equal. 
Similar behavior can be identified in the numerical calculations shown in Fig.~\ref{fig:exp}(e).

In case (ii), shown in Fig.~\ref{fig:exp}(c), the bias field is set to $B^{(\text{ii})}=B_{\text{obs}}(b_0=1.45$G$)$. 
Since now the initial dimer energy is lower, the dressed dimer level crosses the threshold and reaches a separation with the continuum (Fig.~\ref{fig:exp}(a)). 
The oscillations are visible here with two minima and two maxima that can be readily identified, and they agree with a frequency of $\sim 80$kHz supported by the numerical calculations shown in Fig.~\ref{fig:exp}(f).
At longer times the oscillations wash out in both experiment and theory due to the broadband continuum~\cite{Giannakeas19}. 

Case (iii) is shown in Fig.~\ref{fig:exp}(d) for which the bias field is set to $B^{(\text{iii})}=B_{\text{obs}}(b_0=0.77$G$)$. 
As the dressed dimer energy opens a larger gap with the continuum, a faster oscillation frequency is observed.
Also here two minima and two maxima can be identified and their frequency agrees with $\sim 110$kHz.
Fig.~\ref{fig:exp}(g) shows the corresponding numerical calculations and confirms a faster oscillation frequency.
Note, however, that the frequency is not constant. 
Close examination of both experimental and theoretical curves shows that the short time dynamics is somewhat slower. 
This is because for $\tau\lesssim\tau_\text{on}$ the pulse amplitude $b(t)$ does not reach its maximal value $b_0$ and the opened gap is therefore smaller. 

Although numerical calculations qualitatively agree with the experimental measurements, quantitative discrepancies are visible.
The largest discrepancy can be associated with the signals' contrasts.
For example, for incoherent dimer production the theory underestimates the production efficiency by about a factor of two.
It can be partially compensated by the fact that the numerical calculations neglect inelastic losses of dimers. 
While this assumption is correct for all pulse durations used here, in the experiment more time passes between the end of the pulse and the detection during which the majority of dimers are lost.
As the dominant loss mechanism involves inelastic collisions between a dimer and a free atom, the measured number of atoms should be smaller by a factor of $\sim 1.5$.
Although in this case the agreement would improve, for other sets of measurements the oscillation contrast in the numerical calculations are larger by a factor of $\sim 3$ (for the first oscillation) and the inclusion of losses will only increase this discrepancy.

We believe that the main disagreement comes from the fact that the numerical calculations are performed with a zero-range model where the universal dimer energies are matched with those obtained from the experimental measurements.
Although we work with the scattering length an order of magnitude larger than the van der Waals length of the interatomic potential, the details of the true potential are already responsible for deviation of the binding energy from the universal formula~\cite{Gross11,supMat}.
Thus, these details play a role in dimer association especially for short pulse durations ($\sim\mu$s) for which the bandwidth is comparable to the energy gap between the dressed state and the continuum threshold.
Note also that Feshbach resonances in $^7$Li are all narrow in character~\cite{Julienne14} which makes a single channel approach less reliable. 

In conclusion, a robust coherent phenomenon in the conversion efficiency of Feshbach dimers formed from a thermal gas of ultracold atoms has been experimentally demonstrated and compared with theory.
Our proof of concept studies can be improved in many ways.
Access to lower temperatures, limited by technical reasons, should improve the contrast of oscillations and allow the study of decoherence rates in the system.
This, in turn, opens the way to extend the system to more complex scenarios and to address questions regarding the environment influence on the observed coherent phenomenon.
If the density is increased elastic and inelastic collisions between the superposition state constituents might set in and cause collisional broadening or narrowing which can be revealed via measuring decoherence rates as a function of density.
Such adjustability of the continuum is an excellent example for the control the platform of ultracold atoms  can provide for the system under study.

This research was supported in part by the Israel Science Foundation (Grant No. 1543/20) and by a grant from the United States-Israel Binational Science Foundation (NSF-BSF Grant 2019795), Jerusalem, Israel, and the United States National Science Foundation (NSF Awards 1912350 and 2207977).



\begin{thebibliography}{10}
\bibitem{com2}
Note that \cite{Demekhin12a} contains flaws in the theoretical and numerical
  results, see \cite{Baghery17}.
\end{thebibliography}

\clearpage
\newpage

\setcounter{section}{0}
\setcounter{equation}{0}
\setcounter{figure}{0}
\renewcommand\theequation{S\arabic{equation}}
\renewcommand{\figurename}{{\bf Extended Data Fig.}}
\renewcommand\thefigure{{\bf \arabic{figure}}}

\section*{Supplemental Material}

{\bf Experimental Details.}
In the experiment $B_{\text{obs}}$ is measured as a function of current $I$ drawn by the RLC circuit from its power supplier.
Assuming $b=\alpha I$ and fitting the result with Eq.~(2) of the main text we extract the conversion coefficient to be $\alpha_{\text{exp}}=0.472\pm0.006$~G/A. 
The same coefficient can be directly calculated from the Biot-Savart (BS) law and the known geometry of our current coils which gives $\alpha_{\text{BS}}=0.5(1)$~G/A. 
Although the uncertainty on the BS calculation is quite large (due to geometrical uncertainties) the two values are in a very good agreement with each other.

\begin{figure}
	\centering
	\includegraphics[width=0.9\columnwidth]{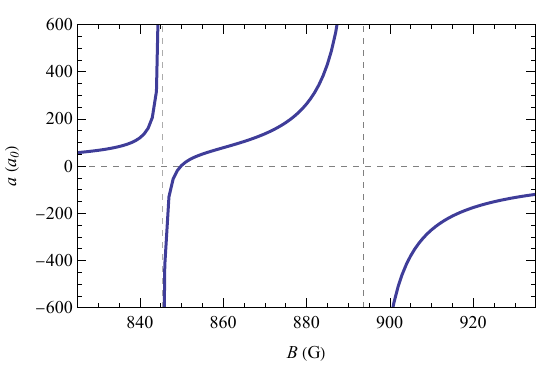}
	\caption{\label{fig:Li7_scat}
		\textbf{Feshbach resonances.}
		Coupled channels calculation data for the scattering length as a function of the magnetic field obtained for $^7$Li $|J=-1/2,m_F=0\rangle$ state. Diverging scattering length indicates positions of two Feshbach resonances at $845.3$ G and $893.78$ G.
	}
\end{figure}

{\bf Notes on the Derivation of Eqs.~(1) $\&$ (2).} 
As specified in the main text, we work with $^7$Li atoms polarized on $|m_J=-1/2,m_F=0\rangle$ internal state.
The state features two $s$-wave Feshbach resonances shown in Extended Data Fig.~\ref{fig:Li7_scat}~\cite{Julienne14}.
We fit this scattering length data with a factorized expression that incudes two poles $B_{\rm res}^{(i)}$ where $i\in\{1,2\}$:
\begin{equation}
a(B)=a_{bg}\left(1- \frac{\Delta_1}{B-B_{\rm res}^{(1)}}\right) \left(1- \frac{\Delta_2}{B-B_{\rm res}^{(2)}}\right),
\label{Eq:SL}
\end{equation}
In Table~\ref{table:FeshbachResonanceParam} the extracted parameters of both resonances are summarized along with differential magnetic moments of the molecular channel with respect to the free-atom continuum $\delta\mu_i$, and the effective ranges $r_e (B=B_{\rm res}^{(i)})$ at the respective resonance positions.
\begin{widetext}

\begin{table}[ht]
\centering
   \begin{tabular}{cccccc}
   \hline\hline
    $B_{\rm res}^{(i)} [G]$ & $\Delta_i$ [G] & $a_{bg} /a_0$ & $\delta \mu_i$ [MHz/G] & $r_e (B=B_{\rm res}^{(i)})/ r_{\rm vdW}$ & s$_{\rm res}$\\
    \hline
     893.78 & -237 & -18.7& -2.66 & $-1.11$ & 0.49\\
     845.3 & 4.64 & -18.7& -2.66 & $-40$ & 0.0467\\
    \hline
   \end{tabular}
   \caption{Parameters characterizing Feshbach resonances. The van der Waals length for $^7$Li atoms is $r_{\rm vdW}\approx 32.6\,a_0$ where $a_0$ is the Bohr radius and $s_{res}$ is the resonance strength~\cite{Chin10}.}
   \label{table:FeshbachResonanceParam}
\end{table}
\end{widetext}

All experiments are performed within the close vicinity ($\leq 0.5$~G) of the magnetic field $B_{\rm bare} = 885$~G as specified in the main text (see Fig.~2 in the main text).
Thus, the following conditions are always well satisfied: $\Delta_1/(B_{\rm bare}  - B_{\rm res}^{(1)})\gg 1$ and $\Delta_2/(B_{\rm bare}  - B_{\rm res}^{(2)})\ll 1$ allowing the simplified expression of Eq.~(\ref{Eq:SL}):
\begin{equation}
a(B=B_{\rm bare})\approx  -\frac{a_{bg}\Delta_1}{B_{\rm bare}-B_{\rm res}^{(1)}}=\left|\frac{a_{bg}\Delta}{B_{\rm bare}-B_{\rm res}}\right| > 0,
\label{Eq:SLPerturbed}
\end{equation}
where we dropped the index (as only the first resonance is relevant) and the minus sign (as $a>0$ throughout this work) for convenience. 
Time modulation of the magnetic field (see Eq.~(3) in the main text) is transferred to the time modulation of the scattering length as follows:
\begin{equation}
\begin{split}
a(t) & \approx  \left|\frac{a_{bg}\Delta}{(B_{\rm bare}+b_0\sin(2\pi\nu t))-B_{\rm res}}\right|\\
&\approx a_{\rm bare}\left(1-b_r\sin(2\pi\nu t)+\mathcal{O}(b_r^2)\right),
\end{split}
\label{Eq:SLPerturbedTimeD}
\end{equation}
where $a_{\rm bare} = a(B=B_{\rm bare})$ and $b_r=\frac{b_0}{|B_{\rm bare}-B_{\rm res}|}\ll 1$ is satisfied quite well even for the largest modulation strength amplitudes. 

\begin{figure}
	\centering
	\includegraphics[width=0.9\columnwidth]{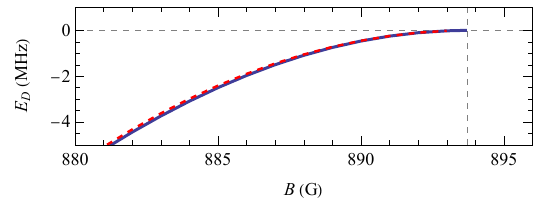}
	\caption{\label{fig:Li7_dimer}
		\textbf{Dimer binding energy.}
		 The coupled channels calculations dimer binding energy in the vicinity of the Feshbach resonance at $893.78$ G is shown as a blue solid line. Dashed-dotted dark red line indicates a dimer whose energy level satisfies a simple expression: $E=-\hbar^2/m (a+r_c)^2$. In the region of interest ($885$~G) this approximation deviates from the exact dimer by $\sim 4\%$.
	}
\end{figure}

The dimer binding energy corresponding to the Feshbach resonance at $893.78$~G is shown in Extended Data Fig.~\ref{fig:Li7_dimer}.
It can easily be observed that in the region of the relevant magnetic field bias the exact dimer is very well approximated by a simple expression: $E=-\hbar^2/m (a+r_{\rm c})^2$ where $r_{\rm c}$ is the correction to the universal $E\propto -1/a^2$ threshold law.
The correction is valid only if $a \gg r_{\rm c}$ and it includes two terms: the finite range of the potential and the resonance strength. 
Its rigorous form is $r_{\rm c} = \bar{a} (s_{res}^{-1} - 1)$ where $\bar{a}=4\pi/\Gamma(1/4)^2\, r_{\rm vdW}$ and $\Gamma(x)$ is the Gamma function~\cite{Chin10}.
The time modulated binding energy is then:
\begin{equation}
E_d(t) = -\frac{\hbar^2}{m (a(t)+r_{\rm c})^2}
\end{equation}
For vanishing modulation strength (see Eq.~\ref{Eq:SLPerturbedTimeD}) we expand this expression in powers of $c_r = r_{\rm c}/a_{\rm bare}$ which in our experiment is $\sim 0.08$ (see the definition of $r_c$ and Table~\ref{table:FeshbachResonanceParam}):
\begin{equation}
E_d \approx -\frac{\hbar^2}{m a_{\rm bare}^2} \left(1-2 c_r+3 c_r^2+\mathcal{O}(c_r^3)\right).
\end{equation}
Returning to time dependence we substitute $a_{\rm bare}=a(t)$ from Eq.~\ref{Eq:SLPerturbedTimeD} in the above expression and expand it in powers of $b_r$.
Keeping in mind that even for largest modulation amplitudes $b_r$ remains of the same order of magnitude as $c_r$, the expansion is kept up to the same order:
\begin{equation}
\begin{split}
E_d(t) & \approx E_{\rm bare}  \left(1-2 c_r+3 c_r^2+\mathcal{O}(c_r^3)\right)\times \\\ & \left( 1 + 2 b_r\sin(2\pi\nu t) + 3 b_r^2 \sin^2(2\pi\nu t) + \mathcal{O}(b_r^3)\right),
\end{split}
\end{equation}
where we introduced $E_{\rm bare}=-\hbar^2/ma_{\rm bare}^2$ as a binding energy for vanishing modulation strength at the universal threshold law limit ($a_{\rm bare}\rightarrow \infty$).  
Taking the time average over an oscillation period and leaving only the leading order non-vanishing correction in $b_r$, we arrive at Eq.~(1) of the main text: 
\begin{equation}
\langle E_d(t) \rangle =E_d(b_0) \approx  A E_{\rm bare} \left(1 + \frac{3}{2 A}b_r^2\right),
\label{Eq:MeanEnergy}
\end{equation}
where $A = 1 - 2 c_r + 3 c_r^2 \approx 0.86$ is the $b_0$-independent constant.

As explained in the main text, in the experiment we scan the magnetic field bias rather than rf frequency. 
It is then more practical to revamp Eq.~\ref{Eq:MeanEnergy} as a shift in the magnetic field bias.
For that we formally represent $\langle E(t)\rangle$ as a function of $B$ (with the help of Eq.~\ref{Eq:SLPerturbed}):
\begin{equation}
\langle E(t)\rangle= -\frac{\hbar^2}{m a(B)^2} = -\frac{\hbar^2}{m}\left(\frac{B-B_{\rm res}}{a_{bg}\Delta}\right)^2,
\end{equation}
and substitute it into Eq.~\ref{Eq:MeanEnergy}  which becomes:
\begin{equation}
(B-B_{\rm res})^2  \approx  A({B_{\rm bare}-B_{\rm res}})^2 \left(1+ \frac{3}{2A}b_r^2\right).
\end{equation}
Taking the square root of the last expression and expanding it to the lowest order in the small parameter $b_r$ yields Eq.~(2) of the main text.
Note that the magnetic field increases with $b_0$ which suits the intuition that the binding energy deepens with the increase of the modulation strength.
This is because the modulation frequency $\nu$ is kept constant, so that we need to increase the magnetic field bias to compensate for deepening of the binding energy when $b_0$ increases. 

{\bf Theoretical Model.}
In the following additional information is provided on the theoretical model which is employed in Fig.~4 of the main text.
In particular we consider two atoms at low collisional energies which interact via a time-dependent Fermi-Huang $s$-wave pseudopotential
\begin{equation}
    V(\boldsymbol{r}, t) = \frac{4 \pi \hbar^2}{m} a(t) \delta(\boldsymbol{r}) \frac{\partial}{\partial r}(r \cdot~~),
    \label{eq:s1}
\end{equation}
where $a(t)$ indicates a time-dependent scattering length which obeys the following relation:
\begin{equation}
    a(t)=a_\textrm{bare}+a_{\rm{max}} \chi(t) \sin(2 \pi \nu t),
    \label{eq:s2}
\end{equation}
where $a_\textrm{bare}$ denotes the scattering length in the absence of radio-frequency magnetic pulse whose amplitude is $a_{\rm{max}}$. $\nu$ refers to the driving frequency of the pulse and $\chi(t)$ is the pulse's envelope as depicted in Fig.~3, i.e. $\chi(t)=b(t)/b_0$ of the main text.

To gain a better physical intuition, semi-analytical expression of the transition probability density to the molecular state from an initial continuum one of energy $E_{c}$ can be derived within the rotating-wave approximation. 
Namely, Eq.~(5) from Ref.~\citesupp{Giannakeas19}  is adapted for zero-range interactions as in Eq.~(\ref{eq:s1}) yielding the following relations:

\begin{widetext}
\begin{subequations}
\label{eq:s3}
\begin{equation}
	|C_b(t)|^2=  \frac{a_{\rm{max}}^2}{4a^2_\textrm{bare}} \frac{ \sin^2(k a_\textrm{bare})}{\mu \pi k a^3_\textrm{ bare}}\bigg|\int^t dt' \chi(t')e^{-\frac{\gamma J(t,t')}{2\hbar}+i\Phi(t')/\hbar } \bigg|^2
	\label{eq:s3a}\end{equation}
with
\begin{equation}
\Phi(t)  	=\int^{t}_0\!\! dt'(E_R^*(t')-E_\mathrm{in}),~~
E_\mathrm{R}^*(t)=E_\textrm{bare}+h \nu+\frac{a_{\rm{max}}^2}{4a^2_\textrm{bare}} \Delta \chi^2(t)\,,
\label{eq:s3b}
\end{equation}
\end{subequations}
\end{widetext}
where  $E_\textrm{bare}$ is the molecular energy in the absence of the time-dependent field and obeys relation, i.e. $E_\textrm{bare}=-\hbar^2 / m a^2_\textrm{bare}$.
$J(t,t')=\int^t_{t'} dt'' \chi^2(t'')$ and $\gamma=\frac{\pi E_\textrm{bare}^2}{h \nu}\sqrt{-1-h \nu/E_\textrm{bare}}$ denotes the depletion of the bound state into the continuum at $E_\textrm{bare}+h \nu$ energies. 
The time-dependent phase $\Phi(t)$ represents the phase accumulation between a continuum state of energy $E_{c}$ and the intensity shifted dressed molecular state $E_R^*(t)$.
Furthermore, according to Eq.~(\ref{eq:s3b}) we observe that $E_R^*(t)$ describes the modification of the dressed molecular state with energy $E_\textrm{bare}+h \nu$ during the pulse.
Indeed, the latter induces a shift $\Delta$ on the dressed bound state which is proportional to the square of the amplitude of the pulse.
This shift for zero-range interactions can be written in a closed form and it obeys the following relation:
\begin{equation}
    \frac{a_{\rm{max}}^2}{4 a^2_\textrm{bare}}\Delta=-\frac{a_{\rm{max}}^2}{a^2_\textrm{bare}}\frac{E^2_\textrm{bare}}{2 h \nu}.
    \label{eq:s4}
\end{equation}
Note that the shift is negative which originates from the level repulsion between the dressed dimer state and the continuum. 
Recall that the dressed molecular state energetically lies slightly above the threshold; thus there is sufficient density of continuum states that shift the dressed state towards, or even below, the threshold.

Indeed, the intensity shifted dressed molecular state in Eq.~(\ref{eq:s3b}) together with Eq.~(\ref{eq:s4}) provide a qualitative physical insight of Fig.~4(a) shown in the main text.
Apart from that, Eq.~(\ref{eq:s3}) addresses also the underlying physics in the case that the strength of the pulse is not sufficiently strong yielding thus a dressed dimer state above the threshold.
This scenario is shown on Fig.~4(b) and (e) in the main text, and in the following we demonstrate the necessary conditions which yield a maximal conversion efficiency.
For this purpose consider a square-like pulse with FWHM equal to the one used in the experiment and assume that the decay of the dimer state is small, i.e. $\gamma\ll1$.
Under these considerations the transition amplitude of Eq.~(\ref{eq:s3}) at the end of the square pulse, i.e. at $t=\tau_0$, yields the following expression:
\begin{equation}
    |C_b(\tau_0)|^2=\frac{a_{\rm{max}}^2}{\pi a^2_\textrm{bare}}\frac{|E_\textrm{bare}|^{3/2}}{E_{c}-E_\textrm{bare}}\frac{4\sqrt{E_{c}}\sin^2(\frac{E_{c}-\varepsilon_R^*}{2\hbar}\tau_0)}{(E_{c}-\varepsilon_R^*)^2}, \label{eq:s5}
\end{equation}
where $\varepsilon_R^*=E_d+h \nu+\Delta a_{\rm{max}}^2/4a^2_\textrm{bare}$ is the intensity shifted dimer energy.

For a thermal gas the energy level statistics fulfill a Maxwell-Boltzmann distribution. 
This implies that there is no preferable initial continuum state, therefore the probability $|C_b(\tau_0)|^2$ in Eq.~(\ref{eq:s5}) needs to be thermally averaged over an ensemble of continuum states.
The thermally averaged $|C_b(\tau_0)|^2$ yields the fraction of the atoms converted into molecules, i.e conversion efficiency:
\begin{equation}
    \frac{2N_m}{N}=2n \lambda_T^3\int_0^\infty dE_{c} e^{-E_{c}/k_BT}|C_b(\tau_0)|^2,
    \label{eq:s6}
\end{equation}
where $N$ ($N_m$) is the number of atoms (molecules), $n$ is the density of the thermal gas at temperature $T$, and $k_B$ is the Boltzmann constant.
$\lambda_T$ is the thermal de Broglie wavelength, i.e. $\lambda_T=\sqrt{h/(\pi m k_BT)}$ with $m$ being the mass of the atoms.

Substituting Eq.~(\ref{eq:s5}) into Eq.~(\ref{eq:s6}) we numerically obtain in the limit of long pulses that the conversion efficiency is maximized when intensity shifted dimer energy $\varepsilon_R^*$ is approximately equal to $k_B T /2$.
\begin{equation}
    E_\textrm{bare}+h \nu+\Delta a_{\rm{max}}^2/4a^2_\textrm{bare}\approx \frac{k_B T}{2}.
    \label{eq:s7}
\end{equation}

Eq.~(\ref{eq:s7}) demonstrates that for weak radio-frequency pulses the atoms are most efficiently converted into molecules as long as the intensity shifted dimer state and the continuum one with energy $E_{c}=k_B T/2$ are quasi-degenerate, as it is depicted in Fig.~4(a) in the main text.

{\bf Likelihood Analysis.}
\begin{figure}
	\centering
	\includegraphics[width=1.00\columnwidth]{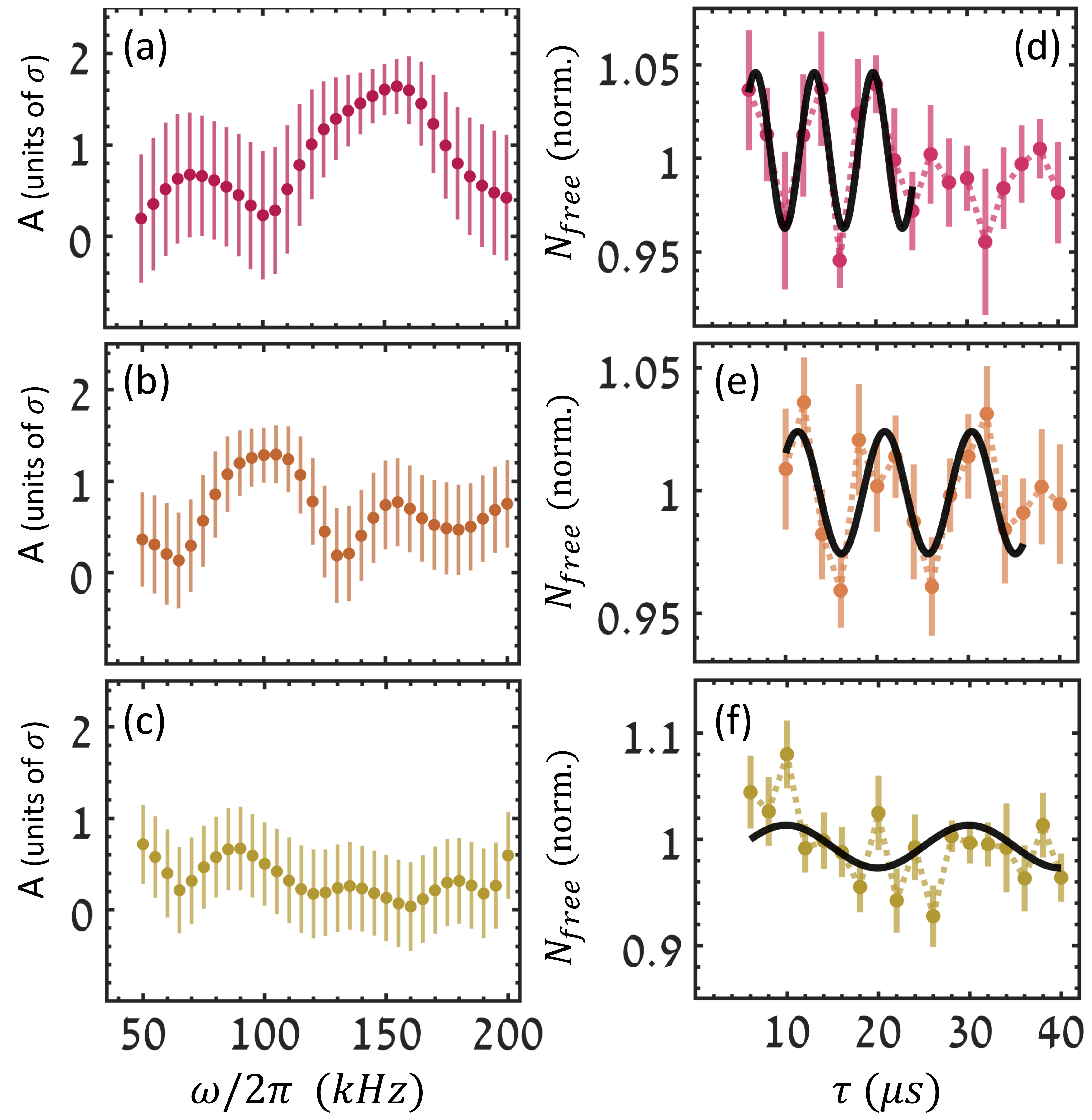}
	\caption{\label{fig:LikelyhoodAnalysis}
		\textbf{Likelyhood analysis.}
		Fourier transform (a,b,c) of the original data limited to initial oscillations (c,d,f). The black solid lines indicate the best fits with $A_1^{max}$ and $A_2^{max}$ (see text). In (f) the black curve indicates the lowest frequency which corresponds to the maximal amplitude.
	}
\end{figure}
\begin{figure}
	\centering
	\includegraphics[width=1.00\columnwidth]{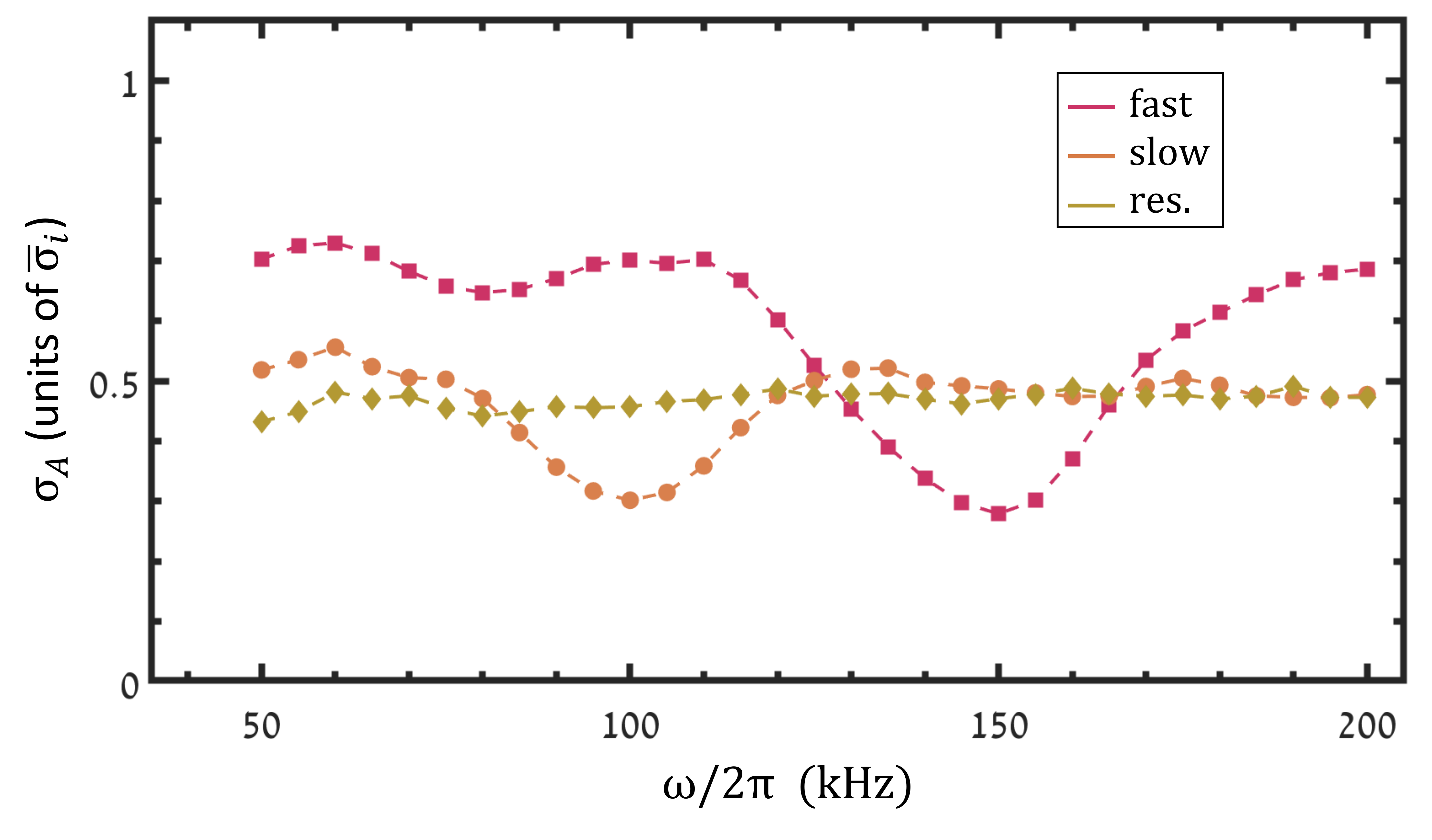}
	\caption{\label{fig:ErrorBars}
		\textbf{Signature of the preferred frequency.}
		Amplitude errors of the Fourier transform presented in Extended Data Fig.~\ref{fig:LikelyhoodAnalysis} (a)-(c). The preferred frequency is manifested by the minimum in the errors with red squares and orange circles representing results of Extended Data Fig.~\ref{fig:LikelyhoodAnalysis} (a),(b) respectively. In contrast, light green diamonds show no pronounced minimum within the relevant spectrum window (Extended Data Fig.~\ref{fig:LikelyhoodAnalysis} (c)) indicating the absence of the preferred frequency.  
	}
\end{figure}
The experimental results shown in this paper suffer from a relatively low signal-to-noise ratio (SNR). 
Here we analyze the likelihood to obtain a similar sequence of data from a sample of random numbers with the same SNR.
Due to a strong decay, there are only two to two and a half oscillations that can be identified in the data.
Therefore, our analysis is concentrated only on this region.
First, we extract the average standard deviations of noise of these oscillations when the mean value is normalized to one.
For two sets of measurements we obtain $\bar{\sigma}_1 = 0.2$ and $\bar{\sigma}_2 = 0.27$ for slow and fast oscillations respectively (see Fig.~4 (c),(d) of the main text).
 We then fit both sequences with $y=y_0+A\sin(\omega t + \phi)$ keeping $y_0, A$ and $\phi$ as fitting parameters and scanning $\omega\in2\pi\times[50,200]$~kHz. 
The resulted dependence of the amplitude $A$ on $\omega$ is shown in Extended Data Fig.~\ref{fig:LikelyhoodAnalysis}(a),(b).  
The maximal amplitudes for both sequences are, then, $A_1^{max}=1.3\bar{\sigma}_1$ and $A_2^{max}=1.65\bar{\sigma}_2$.
The best fits corresponding to $A_1^{max}$ and $A_2^{max}$ are shown in Extended Data Fig.~\ref{fig:LikelyhoodAnalysis}(d),(e).
To show the significance of the maximal amplitude in the previous analysis we apply it to the experimental data for which we claim no preferred frequency (Fig.~4(b)).
However, this data shows strong decay and to remove it we first apply a linear fit to the first $30\,\mu$s and then subtract it from the data.
Next, the analysis proceeds as described above and the results are shown in Extended Data Fig.~\ref{fig:LikelyhoodAnalysis}(c),(f).
To further emphasize the significance of the preferred frequency, the amplitude error of the fit as a function of frequency for all three cases is shown in Extended Data Fig.~\ref{fig:ErrorBars}.
The absence of the preferred frequency in the last case is clearly indicated by the absence of a pronounced minimum in the amplitude error. 
We note that the results shown in Extended Data Fig.~\ref{fig:LikelyhoodAnalysis} are derived from the same two step analysis for all experimental scenarios (i.e. Fig.~4(b),(c),(d)).
However, the linear fit to the data of Fig.~4(c),(d) data makes negligible contribution.

We then generate $10^6$ pairs of sequences of random data with $\sigma_r = \max\{\bar{\sigma}_1,\bar{\sigma}_2\}$ and perform fitting of each sequence with the described above fitting procedure.
The false positive is counted if $A_{max}\geq \min\{A_1^{max},A_2^{max}\}$ for both sequences in a pair and the corresponding frequency of the first falls short of that of the second.
This analysis provides us with less than one false positive out of $10^3$ random sequences which corresponds to a statistical significance of $3.3\,\sigma$ of the reported results. 
This seemingly surprisingly large statistical significance obtained with initially low SNR signals is related to the fact that there are two independent measurements with a meaningful frequency dependence between them.
Note that the actual statistical significance of our results is even better because the above analysis excludes the third signal which, for suitable parameters, do not show oscillations baring an additional piece of meaningful information. 
More fine details of the results that can only strengthen the statistical significance are discussed in the main text.

\end{document}